\documentclass[aps,pra,preprint,superscriptaddress,showpacs,showkeys]{revtex4}
\usepackage{bm}
\usepackage{graphicx}
\begin{document}

\title{Dirac phase and replicating adiabaticity in isotropically moving wall confinement}

\author{Mohammad Mehrafarin}
\email{mehrafar@aut.ac.ir} 
\affiliation{Physics Department, Amirkabir University of Technology, Tehran
15914, Iran}
\author{Reza Torabi} 
\email{rezatorabi@aut.ac.ir}
\affiliation{Physics Department, Tafresh University, Tafresh, Iran}

\begin{abstract}
Geometric phase in the wave function is important with regard to quantum non-locality and  adiabatic evolution. We study the confinement of a particle by three-dimensional isotropically moving walls, of relevance to experimental trapping techniques, via a proposed approach that explains the physical origin of the geometric Dirac phase induced in the wave function. This phase depends only on the relative rate of change of the spatial scale factor. The approach also yields the class of external potentials that replicate adiabatic evolution in finite time. As illustrative examples, we consider uniform and accelerating walls, and the case where the Dirac phase is due to cosmic expansion. 
\end{abstract}

 \maketitle

\maketitle

\section{Introduction}
Dirac showed that when a quantum particle transports in an external
magnetic field, its wave function acquires a phase factor in
addition to the usual dynamic phase \cite{Dirac}. Since its discovery, due to its geometric significance the Dirac phase has appeared in
many other physical situations that involve a $U(1)$ gauge potential \cite{Shapere,Torabi2,Torabi3,Moulopoulos,Savinov,Yang}. Of relevance to this work is the appearance of the Dirac phase in the wave function of particles confined within moving walls. This type of confinement models a quantum piston \cite{Quan} and is of relevance to experimental trapping techniques in condensed matter systems \cite{Meyrath, van Es}.

The earliest article on moving wall confinement was reported by Doescher and Rice \cite{Doescher}. They treated the problem in one dimension by means of a complete set of
functions that were exact solutions of the time-dependent Shr\"{o}dinger equation (TDSE). Their work
sparked interest in further treatment of the problem, using both approximate \cite{Pinder,Mostafazadeh,Munier} and 
exact approaches \cite{Makowski1,Makowski2,Makowski3,Devoto,Dodonov,Morales,Cervero}. Seba \cite{Seba} studied the confinement in connection with the quantum Fermi accelerator model, and the case of time-periodic
wall motion was considered by Scheiniger \cite{Scheiniger}. Pereshogin and Pronin \cite{Pereshogin}
considered the situation from a geometrical point of view using fiber bundle geometry and derived an effective Hamiltonian that was treated by perturbation theory. The appearance of a geometric phase factor in the solution was first pointed out by Greenberger \cite{Greenberger}. This phase factor, which is free from topological complications of the Aharonov-Bohm-type effects, signifies the non-locality of quantum mechanics as the particle can be very far from the boundary. Another significance of the phase factor is in relation with the adiabatic theorem: A system with time-varying Hamiltonian remains in its instantaneous eigenstate, only picking up a geometric phase factor in its wave function in addition to the dynamic phase, provided the time variation is infinitely slow \cite{Berry}.   Moving-wall problems, thus, extend the domain in which the validity of the adiabatic theorem can be examined \cite{Pinder}. This is relevant to implementing shortcuts to adiabaticity, an emergent field that aims at replicating adiabatic state evolution without the requirement of quasi-static time variation \cite{Cirac, Campo1,Campo2,Campo3}.

The literature in the field, such as cited above, do not shed light on understanding why moving walls induce a geometric Dirac phase, and this, to some extend, has hindered due attention to this important factor in moving-wall traps \cite{Mostafazadeh}.  In this work, we consider confinement by three-dimensional isotropically moving walls and derive the solution of the TDSE via a physical approach that explains the origin of the Dirac phase in the wave function. Our approach is based on coordinate transformation, {\it prior} to quantization, to the frame comoving with the wall, which fixes the boundary condition at the expense of introducing a simple $U(1)$ gauge potential. This immediately explains that the appearance of the Dirac phase factor is a direct result of the moving wall. The Dirac phase is independent of the shape of the wall and depends only on the relative rate of change of the spatial scale factor associated with the isotropic motion. Because the comoving frame is non-inertial, a fictitious force emerges whose potential energy enters the transformed TDSE. Using this equation, we deduce the class of external potentials that replicate adiabatic state evolution in finite-time. Such potentials provide shortcuts to adiabaticity in moving wall traps. Thus, the particle remains in the instantaneous energy eigenstate, only picking up a Dirac phase factor in its wave function in addition to the usual dynamic phase, despite non-adiabatic evolution. 

As illustrative examples, we consider the free particle in uniform and accelerating walls, and the case of cosmic expansion, wherein the comoving frame is the cosmic frame carried along with the cosmic expansion, i.e., the frame of the cosmic microwave background radiation and the relative rate of change of the scale factor coincides with the Hubble parameter (see e.g. \cite{Liddle}). On cosmological scales, the universe becomes effectively homogeneous and isotropic with galaxies flying apart from one another. On these scales, the universe expands uniformly under the action of gravity alone and adopting the Newtonian dust model \cite{Liddle}, we derive the gravitational potential energy of a particle (galaxy). Considering the quantum mechanics of such particles, this potential energy is a particular instance of the above mentioned class of potentials.

\section{Dirac phase and the TDSE in comoving coordinates}
We consider the single-particle SE,
\begin{equation}
i\hbar\frac{\partial}{\partial t}\psi({\bm
x},t)=-\frac{\hbar^2}{2M}\nabla^2_{\bm x}\psi({\bm x},t)+V({\bm x},t)\psi({\bm x},t)
\label{one}
\end{equation}
with $\psi= 0$ on isotropically moving boundary wall(s). We denote by, $a(t)$, the scale factor associated with the isotropic motion, such that the enclosed volume instantaneously scales by the factor $a^3(t)$. Examples of this moving boundary condition are the spherical wall $r=a(t)r_0$, and the cubic walls $x_i=\pm a(t)x_0$ ($i=1,2,3$). We can, thus, introduce a new coordinate system in which the coordinates are fixed, while the separation between points in every direction scales by the factor $a(t)$. These comoving coordinates, which we denote by ${\bm X}$, are related to the physical coordinates by ${\bm x}=a{\bm X}$, so that in the comoving frame the boundary condition is time-independent. Since the comoving coordinates are fixed in time, in the comoving frame the velocity associated with each point must vanish. Therefore, because
\begin{equation}
{\dot{\bm x}}={\dot a(t)} {\bm X}=H(t){\bm x}\label{vel}
\end{equation}
where $H={\dot a}/a$, transforming to the comoving frame entails the Galilean transformation
\begin{equation}
{\bm p}\rightarrow {\bm p}+MH{\bm x}
\end{equation}
for the momentum. The above equation pertains to the momentum operator if we transform to the non-inertial comoving frame {\it prior} to quantization. (This is important to note, because if we transformed after quantization, the momentum operator would just rescale by the factor $a^{-1}$.) In this manner, we fix the boundary condition at the expense of introducing
the gauge potential ${\bm A}=-MH{\bm x}$. This implies that wave function in the physical and the comoving coordinate systems are related by the Dirac phase factor $e^{i\gamma}$, where ($r=|{\bm x}|$)
\begin{equation}
\gamma=-\frac{1}{\hbar}\int {\bm A}\cdot d{\bm
x}=\frac{1}{2\hbar}MHr^2.\label{gamma}
\end{equation}
Note that $\gamma$ this is independent of the shape of the wall and depends only on the relative rate of change, $H$, of the scale factor. We, therefore, have
\begin{equation}
\psi({\bm x},t)=[a(t)]^{-3/2} \phi({\bm x}/a,t)e^{i\gamma} \label{psi}
\end{equation}
where $\phi({\bm X},t)$, satisfying the fixed boundary condition, is the wave function in the comoving frame, and the factor $a^{-3/2}$ is introduced to preserve normalization in both coordinate systems:
\begin{equation}
\int |\psi({\bm x},t)|^2 d^3{\bm x}=\int |\phi({\bm X},t)|^2 d^3{\bm X}.
\end{equation}

Furthermore, because the comoving coordinates are fixed in time, the acceleration associated with each point 
\begin{equation}
{\ddot{\bm x}}=(\dot{H}+H^2){\bm x}=\frac{\ddot{a}}{a}{\bm x}\label{accel}
\end{equation}
must also vanish in this frame. This entails the appearance of a fictitious force field $-\frac{\ddot{a}}{a}M{\bm x}$ in the (non-inertial) comoving frame, whose potential energy, $\frac{1}{2}\frac{\ddot{a}}{a}M r^2$, enters the transformed TDSE.  Hence, the TDSE satisfied by $\phi$ is
\begin{equation}
i\hbar\frac{\partial}{\partial t}\phi({\bm x}/a,t)=-\frac{\hbar^2}{2M}\nabla^2_{\bm x}\phi({\bm x}/a,t)+{\bigg [}\frac{1}{2}\frac{\ddot{a}}{a}Mr^2+V({\bm x},t){\bigg ]}\phi({\bm x}/a,t) \label{phi}
\end{equation}
which can be corroborated by direct substitution of the solution (\ref{psi}) into equation (\ref{one}). In terms of ${\bm X}$ and the new time variable
\begin{equation}
\tau(t)= \int^{t}\frac{dt'}{a^2(t')}
\end{equation}
equation (\ref{phi}) reads:
\begin{equation}
i\hbar\frac{\partial}{\partial \tau}\phi({\bm X},\tau)=-\frac{\hbar^2}{2M}\nabla^2_{\bm X}\phi({\bm X},\tau)+{\bigg [}\frac{1}{2}M \alpha |{\bm X}|^2+a^2V(a{\bm X},\tau){\bigg ]}\phi({\bm X},\tau). \label{phi2}
\end{equation}
where $\alpha(\tau)=a^3\ddot{a}$ and $\phi$ vanishes on boundary wall. This is the TDSE in comoving frame, which can be handled in the standard manner owing to its fixed boundary condition. 

\section{Potentials replicating adiabatic state evolution}
TDSE (\ref{phi2}) can be solved via separation of variables when the expression inside the square brackets is a function of ${\bm X}$ alone. This corresponds to the class of external potentials of the form
\begin{equation}
V(\bm{x},t)=\frac{1}{a^2}\tilde{V}(\bm{x}/a)-\frac{1}{2}\frac{\ddot{a}}{a}Mr^2\label{pot}
\end{equation} 
where $\tilde{V}(\bm {x})$ is any arbitrary time-independent potential. The second term on the right hand side coincides with the well known auxiliary potential term that must be supplemented to implement shortcuts to adiabaticity in moving wall traps \cite{Cervero,Campo1,Campo2}. As an interesting subclass, $\tilde{V}$ can be an arbitrary homogeneous function of degree $-2$, in which case the first term on the right hand side becomes $r^{-2}f(\theta,\varphi)$. The intriguing inverse square potential is relevant to the Efimov effect for three-boson interaction first predicted for nucleons \cite{Efimov} and experimentally observed in ultracold atoms \cite{Kraemer}, the interaction of an electron with a polar molecule \cite{Camblong1}, as well as scalar fields in Nordstr\"{o}m \cite{Camblong2} and anti-de Sitter spacetimes \cite{Moroz}.

Thence
\begin{equation}
\phi({\bm X},\tau)=u({\bm X})\exp(-\frac{i}{\hbar}E\tau)\label{sep}
\end{equation}
where $u$, vanishing on boundary wall, satisfies the time-independent SE (TISE) with eigenvalue E, 
\begin{equation}
-\frac{\hbar^2}{2M}\nabla^2_{\bm X}u({\bm X})+\tilde{V}(\bm{X})u({\bm X})=Eu({\bm X}). \label{TISE}
\end{equation}
The complete wave function, as given by (\ref{psi}), is therefore
\begin{equation}
\psi({\bm x},t)=[a(t)]^{-3/2} u({\bm x}/a)\exp{{\bigg (}-\frac{i}{\hbar}\int^{t}E(t')dt'+{i\gamma}{\bigg )}}\label{psi2}
\end{equation}
where $E(t)=E/a^2(t)$. Thus, the particle remains in the instantaneous energy eigenstate $E(t)$, only picking up a Dirac phase factor $\gamma$ given by equation (\ref{gamma}), in addition to the usual dynamic phase. The state (\ref{psi2}), attainable ideally in the limit of adiabatic evolution \cite{Berry}, is hereby exactly obtained in finite-time, thus providing a shortcut to adiabaticity. 

Let us consider a few examples of the potential form (\ref{pot}) below.

\subsection{Free particle in a uniformly moving spherical wall} 
Here, $\psi=0$ for $r=R(t)$, where $R(t)=a(t)r_0$ is the instantaneous radius and $\ddot{a}=0$. Hence, $\tilde{V}=0$ and the boundary condition is $u=0$ for $|{\bm X}|=r_0$. Thus, (\ref{TISE}) yields the normalized solution 
\begin{equation}
u_{nlm}(|{\bm X}|,\theta,\varphi)=\sqrt{\frac{2}{r_0^3}}\frac{1}{ j_{l+1}(k_{nl})} j_l( k_{nl}|{\bm X}|/r_0)Y_{lm}(\theta,\varphi)
\end{equation}
with energy eigenvalues, 
\begin{equation}
E_{nl}=\frac{\hbar^2 k_{nl}^2}{2Mr_0^2}
\end{equation}
where $k_{nl}$ is the $n$th zero of the spherical Bessel function $j_l$, and $|m|\le l=0,1,2,\ldots$, of course. The wave function of the particle, therefore, is 
\begin{equation}
\psi_{nlm}(r,\theta,\varphi,t)=\sqrt{\frac{2}{R^3(t)}}\frac{1}{ j_{l+1}(k_{nl})} j_l( k_{nl}r/R(t))Y_{lm}(\theta,\varphi)\exp{\bigg (}-\frac{i}{\hbar}\int^tE_{nl}(t')dt'+i\gamma {\bigg )}
\end{equation}
where  
\begin{equation}
E_{nl}(t)=\frac{\hbar^2 k_{nl}^2}{2MR^2(t)}
\end{equation}
are the instantaneous energy eigenvalues.

\subsection{Free particle in a cubic box with uniformly moving walls}
At time $t$, the box has sides $L(t)=a(t)x_{0}$, where $\ddot{a}=0$. We take the origin at the center of the cube, which will remain so at at all times. The boundary conditions are, thus, $\psi=0$ for $x_i=\pm L(t)/2$, so that, $u=0$ for $X_i=\pm x_0/2$. Setting $\tilde{V}=0$ in (\ref{TISE}) yields the normalized solutions  
\begin{equation}
u_{nlm}({\bm X})={\sqrt\frac{8}{x_0^3}}\ f {\bigg(}\frac{n\pi X_1}{x_{0}}{\bigg)}f {\bigg(}\frac{l\pi X_2}{x_{0}}{\bigg)}f {\bigg(}\frac{m\pi X_3}{x_{0}}{\bigg)}
\end{equation}
where $f(x)$ stands for $\sin x$ or $\cos x$
and $n,l,m$ are restricted to even (odd) integer values for $\sin$ ($\cos$). These eight solutions have definite parity, four of which are even and the other four are odd. The corresponding energy eigenvalues are
\begin{equation}
E_{nlm}=\frac{\hbar^2 \pi^2}{2Mx_0^2}(n^2+l^2+m^2).
\end{equation}
Therefore, the wave function of the particle is given by
\begin{equation}
\psi_{nlm}({\bm x},t)={\sqrt\frac{8}{V(t)}}f {\bigg(}\frac{n\pi x_1}{L(t)}{\bigg)}f  {\bigg(}\frac{l\pi x_2}{L(t)}{\bigg)}f {\bigg(}\frac{m\pi x_3}{L(t)}{\bigg)}\exp{\bigg (}-\frac{i}{\hbar}\int^tE_{nlm}(t')dt'+i\gamma {\bigg )}\label{*}
\end{equation}
where, $V(t)=L^3(t)$, is the instantaneous volume of the box and
\begin{equation}
E_{nlm}(t)=\frac{\hbar^2 \pi^2}{2ML^2(t)}(n^2+l^2+m^2)
\end{equation}
are the instantaneous energy eigenvalues. Equation (\ref{*}) reconciles with and generalizes the result obtained in one dimension in reference \cite{Cervero}. As noted previously, $\gamma$ is independent of the wall geometry: The Dirac phase here is identical to that appearing in the case of the spherical wall (i), provided $H$ is the same.

\subsection{Free particle in an accelerating spherical wall}
The boundary conditions are those of  case (i), but $\ddot{a}>0$ and $\tilde{V}(\bm{X})=\frac{1}{2}M\alpha |\bm{X}|^2$. The particle is no longer free in the comoving frame, it is affected by the fictitious acceleration. For constant $\alpha$, equation (\ref{TISE}) pertains to a three-dimensional simple harmonic oscillator contained within the fixed spherical wall. Writing $\alpha=\omega^2$, the solution in spherical coordinates is
\begin{equation}
{\bigg (}\frac{M\omega}{\hbar}{\bigg )}^{\frac{2l+3}{4}}|{\bm X}|^l\exp{\bigg(}{-\frac{M\omega}{2\hbar}|{\bm X}|^2}{\bigg )} L_{(n-l)/2}^{l+1/2}{\bigg(}{\frac{M\omega}{\hbar}|{\bm X}|^2}{\bigg )} Y_{lm}(\theta,\varphi)
\end{equation}
where $L$ denotes the associated Laguerre polynomial,  $l=0,1,2,\ldots$; $n=l, l+2, l+4, \dots$; and $|m|\le l$. The corresponding energy eigenvalues are $(n+\frac{3}{2})\hbar \omega.$
$u$ must vanish for $|{\bm X}|=r_0$, which implies that $n\ne l$ (as $L_0^q=1$), and
\begin{equation}
\omega_{nls}=\frac{\hbar}{Mr_0^2} k_{snl}
\end{equation}
where $k_{snl}$, $s=1,2,\ldots,\frac{n-l}{2}$, is the $s$th zero of $ L_{(n-l)/2}^{l+1/2}$. Hence
\begin{equation}
u_{nlms}(|{\bm X}|,\theta,\varphi)={\bigg (}\frac{k_{snl}}{r_0^2} {\bigg )}^{\frac{2l+3}{4}}|{\bm X}|^l\exp{\bigg(}{-\frac{k_{snl}}{2r_0^2}|{\bm X}|^2}{\bigg )} L_{(n-l)/2}^{l+1/2}{\bigg(}{\frac{k_{snl}}{r_0^2}|{\bm X}|^2}{\bigg )}  Y_{lm}(\theta,\varphi) 
\end{equation} 
with energy eigenvalues 
\begin{equation}
E_{nls}=(n+\frac{3}{2})\frac{\hbar^2 k_{snl}}{Mr_0^2}.
\end{equation}
Therefore, the wave function of the particle is ($n\ne 0$)
\begin{eqnarray}
\psi_{nlms}(r,\theta,\varphi,t)={\bigg [}\frac{k_{snl}}{R^2(t)} {\bigg ]}^{\frac{2l+3}{4}}r^l\exp{\bigg (}{-\frac{k_{snl}}{2R^2(t)}}r^2{\bigg )} L_{(n-l)/2}^{l+1/2}{\bigg(}{\frac{k_{snl}}{R^2(t)}}r^2{\bigg )} \nonumber \\ \times Y_{lm}(\theta,\varphi)
 \exp{\bigg (}-\frac{i}{\hbar}\int^tE_{nls}(t')dt'+i\gamma {\bigg )}
\end{eqnarray}
where
\begin{equation}
E_{nls}(t)=(n+\frac{3}{2})\frac{\hbar^2 k_{snl}}{MR^2(t)}.
\end{equation}

\subsection {Cosmic expansion}
If we go to large enough scales, in practice tens of mega parsecs, the universe becomes effectively homogeneous and isotropic with galaxies
flying apart from one another. On these cosmological scales, the
cosmological principle applies and the universe, which resembles a fluid whose particles are the galaxies, expands uniformly
under the influence of gravity alone. In the Newtonian dust model for the universe, this fluid is taken to be a pressure-less Newtonian fluid \cite{Liddle}. Denoting the cosmic comoving
coordinates, which are carried along with the cosmic expansion, by ${\bf
X}$, the physical coordinates of a particle (galaxy) can be written as ${\bm x}=a(t){\bm X}$.  The comoving coordinates are fixed in time and the scale factor $a(t)$ measures the universal expansion rate. Equation (\ref{accel}), thus, holds for the acceleration of a particle, where $H$ is called the Hubble parameter. The potential field of the gravitational acceleration causing the expansion is, therefore,  $-\frac{1}{2}\frac{\ddot{a}}{a}r^2$.
On cosmological scales, this the only potential influencing the particles. Using this potential, the gravitational Poisson equation, and the continuity equation for the fluid, one reproduces the  (matter dominated) Friedmann equation of general relativity for the scale factor \cite{Liddle}. Therefore, contrary to the previous examples, it is not at our disposal to set the scale factor arbitrarily. Considering the quantum mechanics of such particles, the potential energy in TDSE (\ref{one}) is, thus, given by  $-\frac{1}{2}\frac{\ddot{a}}{a}Mr^2$, a particular case of the  potential form (\ref{pot}) with $\tilde{V}=0$. Hence, the solution (\ref{psi2}) applies, where now $u$ obeys the free-particle TISE. 
The particle, although being influenced by the gravitational potential, remains in the instantaneous free-particle eigenstate, only acquiring an additional Dirac phase factor due to expansion that depends on the Hubble parameter. 

The Friedmann equation has three independent solutions, one of which has a closed form, namely, $a=(t/t_0)^{2/3}$ (subscript zero denotes present value), corresponding to the flat universe of general relativity. For this case $\gamma=Mr^2/3\hbar t$, which, contrary to the dynamic phase, decreases with epoch. The wave function at the present epoch is, then, given by
\begin{equation}
\psi({\bm x},t_0)=u({\bm x})\exp{\bigg[}{\frac{i}{\hbar}{\bigg(}\frac{Mr^2}{3t_0}-3Et_0{\bigg )}{\bigg]}}.
\end{equation}
The photons of the cosmic background radiation experience the expansion due to gravity just as massive particles do (after all they constitute the physical cosmic comoving frame). 
However, they are not described by the above wave function, of course. These photons reach us from all directions and it would be interesting to find their geometric phase, which could be detectable in properly designed interference experiments.

\section{Conclusion}
Geometric phase in the wave function is important with regard to quantum non-locality and adiabatic state evolution. We have studied the the quantum mechanics of a particle confined within isotropically moving walls by a proposed approach based on coordinate transformation to the comoving frame, which fixes the boundary condition at the expense of introducing a simple $U(1)$ gauge potential. Our approach has two merits: (I) It explains the origin of the geometric Dirac phase in the the wave function. This phase factor is independent of the shape of the wall and depends only on the relative rate of change, $H$, of the scale factor associated with the isotropic motion. (II) It yields the class of external potentials that replicate adiabatic state evolution in finite time, thus providing shortcuts to adiabaticity. As illustrative examples, we have considered the free particle in uniform and accelerating walls, and also  the case of cosmic expansion, wherein the comoving frame is the cosmic frame carried along with the cosmic expansion and, $H$, coincides with the Hubble parameter.

\end{document}